\documentclass[prl,showpacs,showkeys,floats,nofootinbib,a4paper]{revtex4}

\usepackage{amsmath,amssymb}
\usepackage{graphicx}
\usepackage{bm}
\usepackage{xspace}

\newcommand{\be}{\begin{equation}}
\newcommand{\ee}{\end{equation}}

\def\Dzero {D\O\xspace}
\def\Bc {B_c}
\def\jpsi {J/\psi}
\def\electron {e^-}

\begin{document}

\title{Survival of $B_c$ mesons in a hot plasma within a potential model}

\author{W.M. Alberico}
\affiliation{Dipartimento di Fisica dell'Universit\`a di Torino and \\ 
  Istituto Nazionale di Fisica Nucleare, Sezione di Torino, \\ 
  via P.Giuria 1, I-10125 Torino, Italy}

\author{S. Carignano}
\affiliation{Department of Physics, University of Texas at El Paso, El Paso, TX 
79968, USA}

  \author{P. Czerski}
  \email{piotr.czerski@ifj.edu.pl}
\affiliation{The H. Niewodnicza\'nski Institute of Nuclear Physics, Polish 
Academy of Sciences,\\
     ul. Radzikowskiego 152, PL-31-342 Krak\'ow, Poland}
  
  \author{A. De Pace}
\affiliation{Istituto Nazionale di Fisica Nucleare, Sezione di Torino, \\ 
  via P.Giuria 1, I-10125 Torino, Italy}
  
  \author{M. Nardi}
\affiliation{Istituto Nazionale di Fisica Nucleare, Sezione di Torino, \\ 
  via P.Giuria 1, I-10125 Torino, Italy}
  
  \author{C. Ratti}
\affiliation{Dipartimento di Fisica dell'Universit\`a di Torino, \\ 
  via P.Giuria 1, I-10125 Torino, Italy}

\begin{abstract}{We extend a previous work on the study of heavy charmonia and 
bottomonia in a deconfined quark-gluon plasma
by considering the $B_c$ family of mesons. With the introduction of this bound 
state of a charm and a beauty quark, we 
investigate at finite temperature the behavior of the quarkonium, in an energy 
region between the $\psi$ and the $\Upsilon$ states.}
\end{abstract}



\maketitle

\section{Introduction}

In the context of relativistic heavy ion collisions the early production of 
heavy quarkonia and the modulation of their survival 
while crossing the deconfined medium, created and thermalized afterward in the 
collision, has been considered as a meaningful test 
of deconfinement since long ago. The attention has been first focused on 
charmonia and bottomonia since in a single hard collision 
(already in p-p scattering) the relevant pairs $c\bar c$ and $b\bar b$ can be 
produced and can form one meson in the $J/\psi$ or 
$\Upsilon$ families (for recent reviews see, e.g., Refs. \cite{Satz,Satz2}). 
Instead the production of a $B_c$ meson ($c\bar b$ or $\bar c b$) requires, in 
the p-p case, the simultaneous 
occurrence of two hard partonic scatterings, which appears to be less favorable.

On the contrary, the $B_c$ formation could be favored in a nucleus-nucleus 
collision where many more partonic (hard) scatterings can 
occur simultaneously. Hence the $B_c$ production might be significantly enhanced 
in A-A collisions with respect to p-p ones.

{
The number of $B_c$ states coming out from the global process is strongly 
affected, on the one side, by the melting of the initially formed $c\bar{c}$ and 
$b\bar{b}$ 
mesons inside the hot, deconfined plasma; on the other side it is affected by 
the regeneration process (as foreseen, for example, 
within a coalescence model) which, according to several authors \cite{greiner}, 
can be important for hidden-charm and -beauty mesons, while it will certainly
dominate over the initial hard production for the $B_c$ case.}

Here we will mainly focus on the modifications of the binding energy of a  $B_c$ 
meson due to the increasing temperature of the plasma, 
above the critical temperature $T_c$ for deconfinement.

The mass and lifetime of the $B_c$ produced in $p\bar p$ have been measured by 
the CDF and \Dzero collaborations at Tevatron 
by studying both its semileptonic $\Bc\to\jpsi~\electron\nu X$ and the hadronic 
$\Bc\to\jpsi~\pi$ decay channels~\cite{cdf,D0}. 
Being the heaviest of the $B$ family, the $B_c$ meson has not been observed 
until recently, in relatively clean experimental conditions. 
In 2012 the CMS collaboration observed these $\Bc^+$ decays at the LHC a mass of 
$m_{\Bc} = 6.272 \pm 0.003$ GeV for the 
$J/\Psi + \pi^+$ channel and of $m_{\Bc} = 6.265 \pm 0.004$ GeV for the $J/\Psi 
+ 3\pi$ channel~\cite{CMS:2012}.

In order to study the temperature evolution of the mass and energy eigenvalues 
of the $B_c$ mesons, we shall employ a non-relativistic 
potential model, which is justified by the large mass of the constituents quarks 
(see e.g. Refs.\cite{Mocsy09,Bazavov09}). 
In Section 2 we shall shortly review our potential model approach, which is 
based on accurate fits of lattice data for the 
thermodynamical free energy. Then we report our estimates for the dissociation 
temperatures of the $B_c$ family. 
A short discussion of the $B_c$'s decay modes and our conclusions are contained 
in the last Section.

\section{A review of our potential model}

\begin{figure}[h!]
\begin{center}
\includegraphics[clip,width=0.8\textwidth]{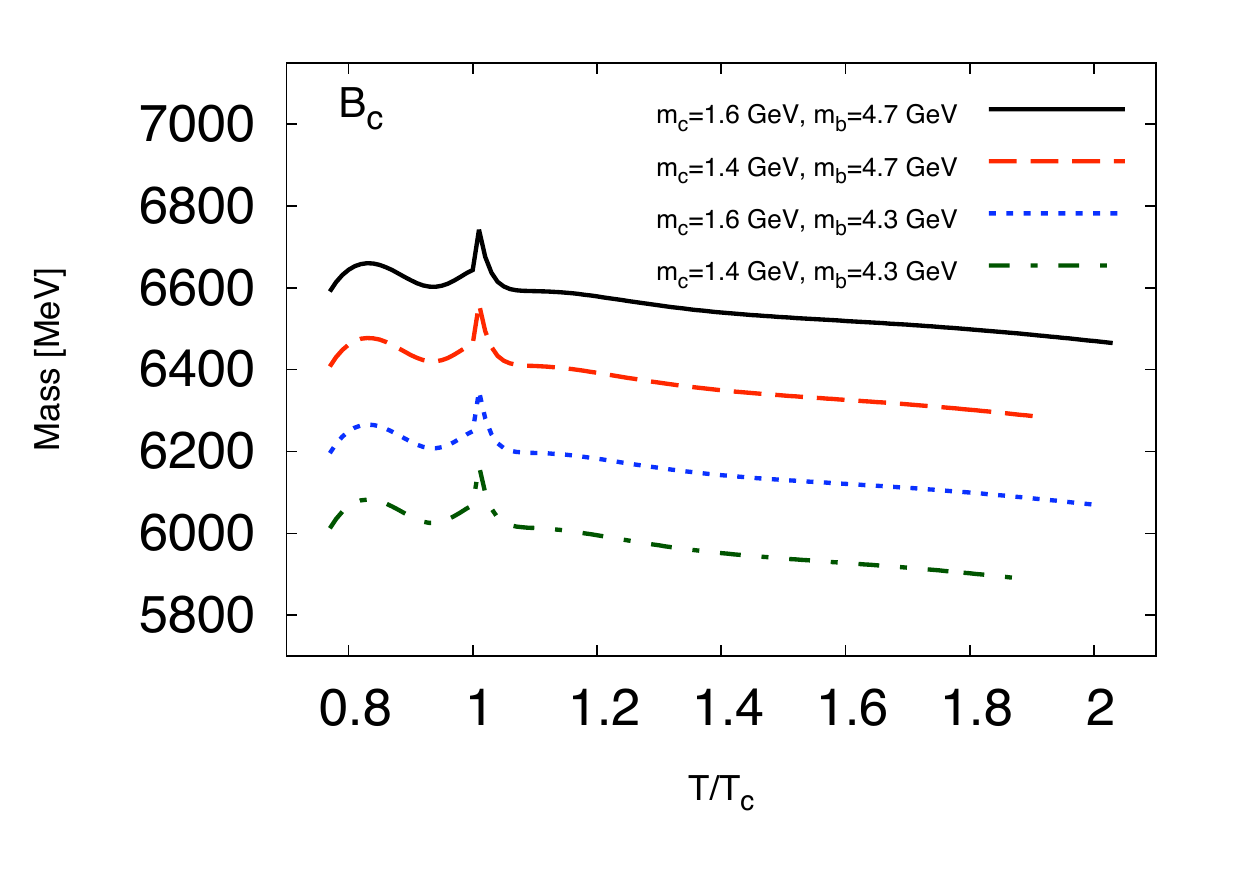}
\caption{Mass as a function of temperature of the lowest $S$-wave $b\bar{c} 
(c\bar{b})$
  state obtained from the solution of the Schroedinger equation. Different 
combinations of constituent quark masses have been investigated.}
\label{fig:massebc}
\end{center}
\end{figure}

The considerable mass of the $B_c$ meson allows us to apply to the $B_c$'s the 
same potential model at finite temperature we previously
developed~\cite{alb1,alb2} in order to study the $J\psi$ and $\Upsilon$ states.
The main points of our approach are the following ones: we suppose that the 
heavy meson bound states are formed in the initial stage 
of the heavy ion collision. These states then cross the deconfined plasma, which 
weakens their binding and in many cases leads to their
dissociation. We chose to encode this behavior into an effective 
temperature-dependent potential, which can be extracted from a fit to lattice 
QCD 
calculations of the color singlet free energy\footnote{The potential model 
adopted here was fitted to lattice data obtained in the approximation of $N_f=2$ 
light flavors \cite{kac05}. We are aware of more recent data 
\cite{Bazavov:2012}, which have been qualitatively improved by increasing the 
number of lattice points for small T. This is however not too relevant for our 
case, since the present approach is valid only close to $T_c$ or above it.}
$F_1$. 
From an accurate series of fits on $F_1$ \cite{alb1}, 
\begin{equation}
 F_1(r,T)=-\frac{4}{3}\frac{\alpha(r,T)}{r}e^{M(T)r}+C(T)
\label{F1}
\end{equation}
 the singlet internal energy was calculated as $U=-T^2\partial(F/T)/\partial T$ 
and the effective heavy quark potential 
$V(r,T)$ was singled out from the medium contributions \cite{wong,alb2}:
\begin{equation}
 V_1(r,T)=U_1(r,T)-U_1(r\to\infty,T).
\end{equation}
In Eq.(\ref{F1}) the coupling $\alpha$ is fixed by the customary RGE, but 
employing a temperature dependent scale, with 
coefficients determined, at each temperature, by the above mentioned fitting 
procedure. For further details we refer the 
reader to Ref.\cite{alb1}.

We then solve the Schroedinger equation associated to the quarkonium states,
\be
 \left(-\frac{\hbar^2}{2\mu}\nabla^2 + V(r,T)\right)\psi({\vec r},T) = 
\epsilon(T)\psi({\vec r},T)\,,
\ee
and obtain the temperature dependent energy eigenvalues, which allow us to 
follow the evolution of the bound state masses and 
to determine the dissociation temperatures, namely the temperature at which the 
binding vanishes.

By evaluating the quarkonium radial wave functions $R(r)$ in the origin, we are 
also able (via a non-relativistic QCD expansion) 
to construct the corresponding spectral functions \cite{bodw,mocsy}, which 
encode many physical properties of these states and allow 
a direct comparison with lattice QCD results \cite{alb2,mocsy}.

\section{Results and discussion.}

We present our results starting with the temperature dependence of the masses of 
the $B_c$ states. The fundamental state ($B_c$), 
the first radially excited state ($B_c'$) and the first P-wave state 
($\chi_{B_c}$) have been considered. 
Our model does not distinguish between states with intrinsically different 
angular momentum, hence the $\Bc$ and its hyperfine partner 
$\Bc *$ appear degenerate.
In order to establish the model dependence on the effective masses chosen for 
the heavy quarks, we performed calculations 
using both $m_c=1.4$ and $1.6$ GeV for the charm quark and both $m_b=4.3$ and 
$4.7$ GeV for the beauty.
 Dynamical thermal masses, defined as the sum of the quark mass and the 
asymptotic value of the $Q\bar Q$ potential, are also 
 introduced \cite{alb2}. While the necessity of including such contributions may 
be questionable, we remark that the resulting 
 variation in the meson masses is almost negligible, being always below 50 MeV 
for $T < T_c$ and below 20 MeV for $T > T_c$.
 \begin{figure}[h!]
 \begin{minipage}{.45\textwidth}
  \includegraphics[clip,width=\textwidth]{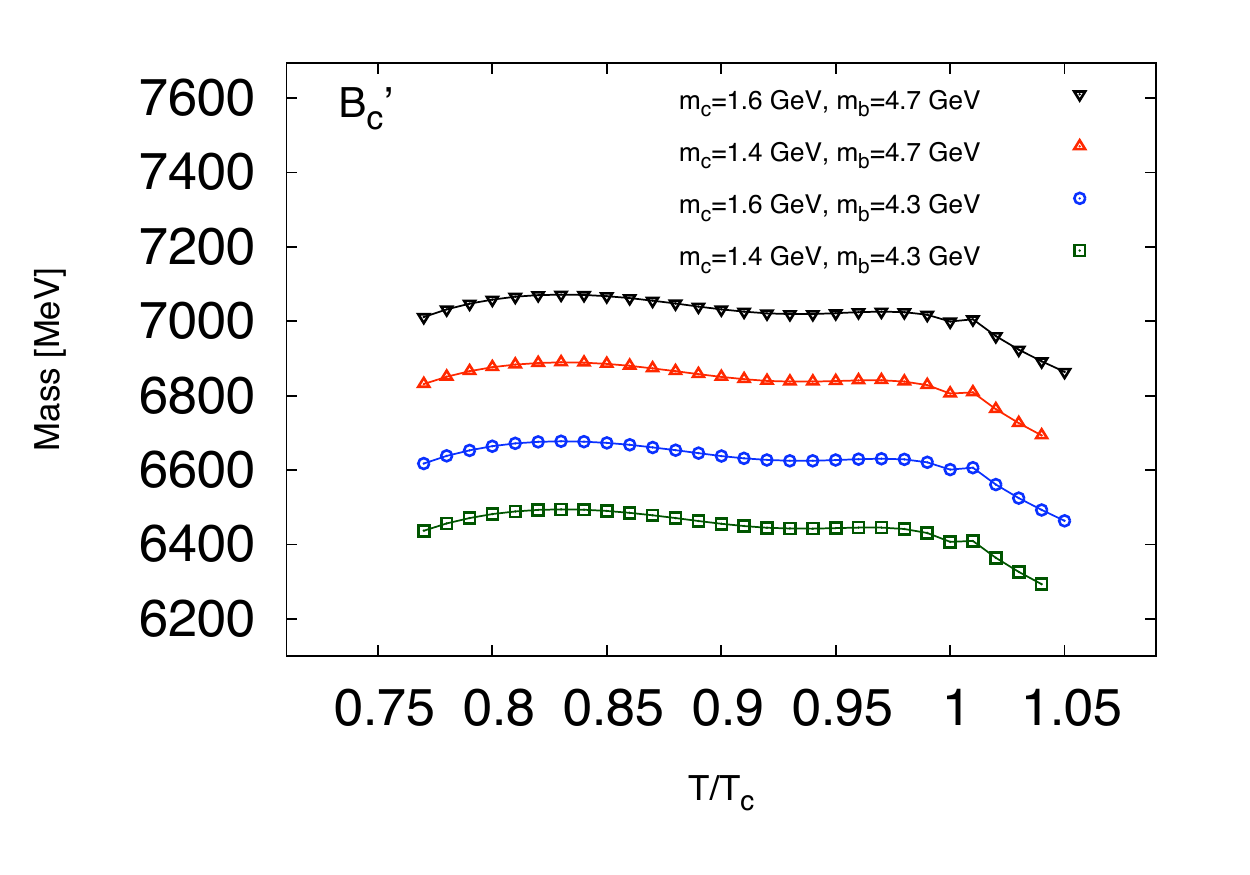}
  \caption{Mass as a function of temperature of the first $S$-wave $b\bar{c} 
(c\bar{b})$ excited
  state.}
\label{fig:massebcprimo}
 \end{minipage}
 \begin{minipage}{.45\textwidth}
  \includegraphics[clip,width=\textwidth]{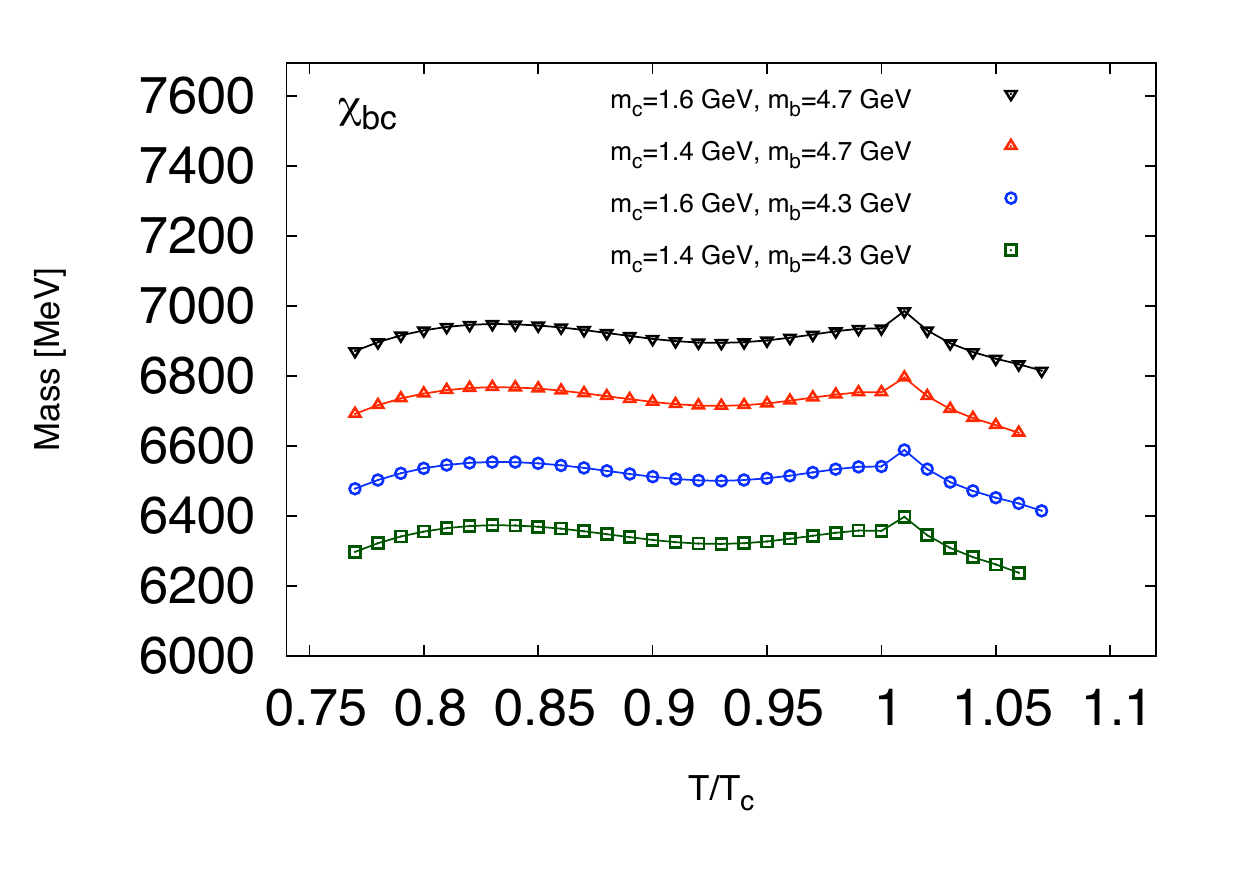}
 \caption{Mass as a function of temperature of the lowest $P$-wave $b\bar{c} 
(c\bar{b})$
  state.}
\label{fig:massechibc}
 \end{minipage}
\end{figure}

The results for the masses of the pseudoscalar ground state $B_c$, for the 
radial excitation $B_c'$ and for the P-wave state $\chi_{B_c}$ 
as a function of $T/T_c$ are shown in figures \ref{fig:massebc}, 
\ref{fig:massebcprimo} and \ref{fig:massechibc}, respectively. 
We recall that perturbative calculations up to order $\alpha^4_s$ for the $B_c$ 
mass give a value of $m_{B_c}=6326^{+29}_{- 9}$~MeV \cite{bramb}, 
while lattice QCD predicts $m_{B_c}=6304\pm12^{+18}_{- 0}$ MeV \cite{lqcdbc} and 
a study with zero temperature potential models 
gives $m_{B_c}=6258\pm20$~MeV \cite{eicht}, the latter being closer to the most 
recent measurement~\cite{CMS:2012}. 
These numbers are quoted for reference only, but cannot be directly compared 
with the mass values 
obtained at finite temperatures: in particular the present approach cannot be 
extended to temperatures much below the critical one.

Table \ref{tab1} shows the dissociation temperatures (defined as the value where 
the binding energy vanishes) obtained for the various 
states, in units of the critical temperature $T_c=202$~MeV (we use here the same 
value of $T_c$ employed in Refs.~\cite{alb1} and \cite{kac05}
for $n=2$ flavors).

\begin{table}
\caption{The dissociation temperatures obtained for the various states, in units 
of the critical temperature $T_c=202$~MeV.\label{tab1}}
\begin{tabular}{lcccc}
\hline
$c\bar b$ $b\bar c$ \quad & \quad $m_c$~=~1.4~GeV \quad & \quad $m_c$~=~1.4~GeV 
\quad & \quad $m_c$~=~1.6~GeV \quad & \quad $m_c$~=~1.6~GeV \\ 
                          & \quad $m_b$~=~4.3~GeV \quad & \quad $m_b$~=~4.7~GeV 
\quad & \quad $m_b$~=~4.3~GeV \quad & \quad $m_b$~=~4.7~GeV \\ 
\hline \hline
$B_c$            & 1.87     & 1.90   & 1.99   & 2.02     \\
    $\chi_{B_{c}}$    & 1.05     & 1.05   & 1.06   & 1.06     \\
    $B_c '$          & 1.03     & 1.04   & 1.04   & 1.05     \\
\hline
\end{tabular}
\end{table}

\begin{figure}[h!]
 \begin{minipage}{.45\textwidth}
  \includegraphics[clip,width=\textwidth]{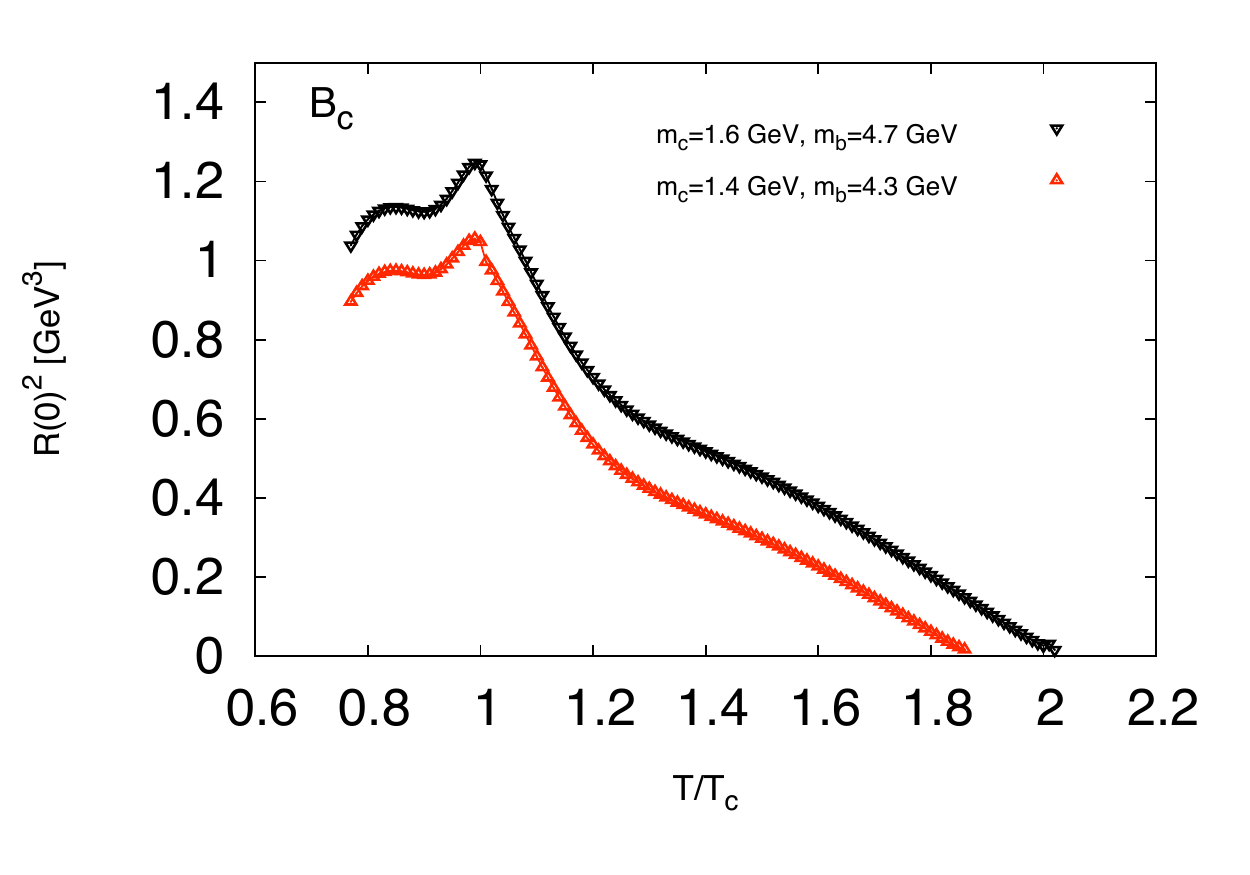}
\caption{Squared value in the origin, for the $b\bar{c}$ system
  ($m_c=1.6$~GeV, $m_b=4.7$~GeV), of the $S$-wave radial wave function as a 
   function of temperature }
\label{fig:r0bc}
 \end{minipage}
 \begin{minipage}{.45\textwidth}
  \includegraphics[clip,width=\textwidth]{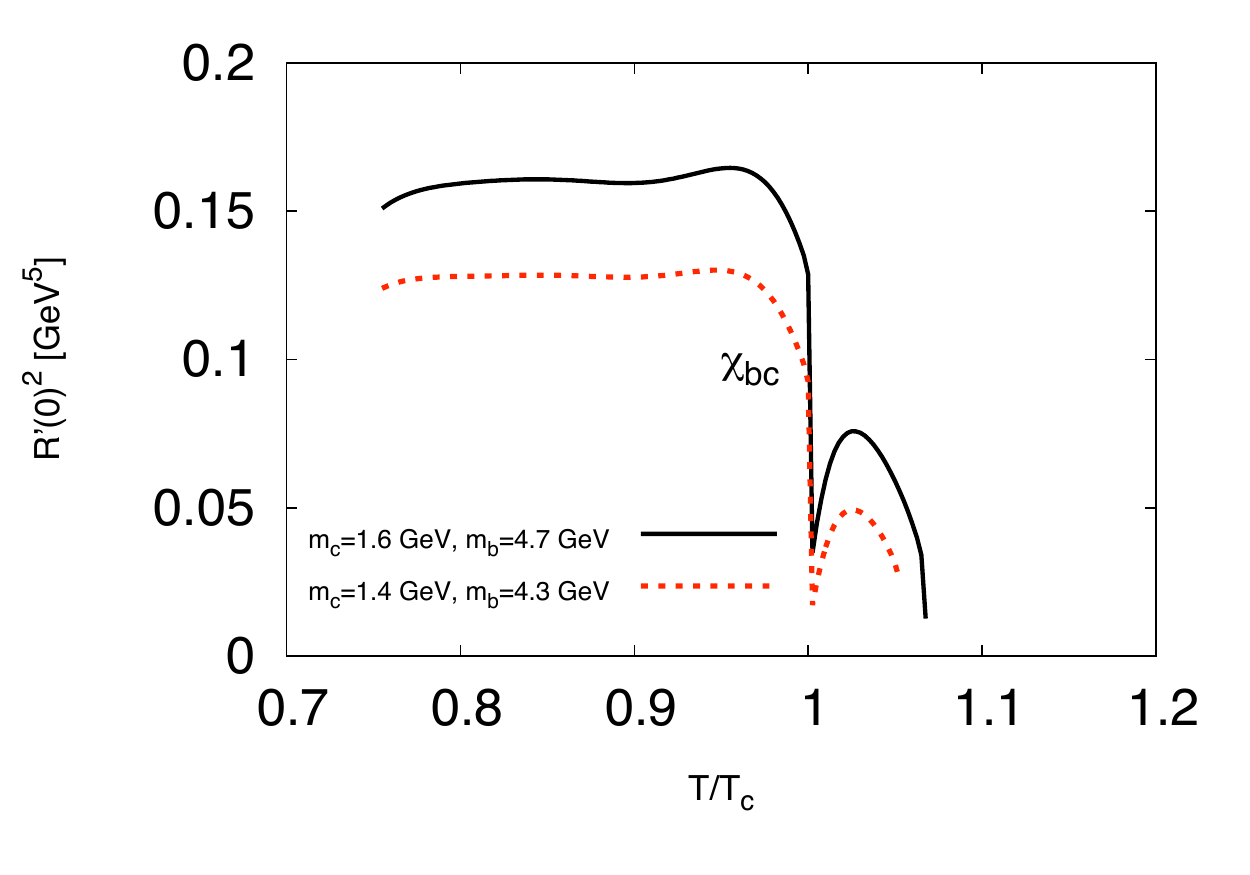}
 \caption{Squared value in the origin for the $b\bar{c}$ of the first derivative 
of the $P$-wave radial wave function, as a function of temperature}
\label{fig:rprimo0chibc}
 \end{minipage}
\end{figure}

We also report in Figs. \ref{fig:r0bc} and \ref{fig:rprimo0chibc} the variation 
with the temperature of the radial wave function (or of its 
first derivative for the P wave state) evaluated in the origin for the $B_c$ and 
$\chi_{B_c}$ states respectively. 
The values obtained by zero temperature potential models \cite{eicht} are 
$|R(0)|^2=1.68$ for the $S$-wave and 
$|R'(0)|^2=0.20$ for the $P$-wave state.

The values of $R(0)$ and $R'(0)$ are then used to build the spectral functions 
at different temperatures. 
Indeed, we recall that the spectral function for a generic meson channel 
$\sigma_M(\omega, T)$ can be written as \cite{alb2,bodw}
\be
 \sigma_M(\omega, T) = \sum_n |\langle 0\mid j_M\mid n\rangle|^2
 \delta(\omega - E_n) = \sum_n F_{M,n}^2\delta(\omega - E_n) + \theta(\omega 
-s_0)F_{M,\epsilon}^2 \,,
\ee
where, for instance, $F_{PS}^2 = \frac{N_c}{2\pi}|R(0)|^2$ for the pseudo-scalar 
state and
 $F_{S}^2 = \frac{9N_c}{2\pi m^2}|R'(0)|^2$ for the P-wave scalar state 
\cite{bodw}.

In order to achieve a better comparison with lattice and perturbative QCD 
results, we add, respectively, a finite width to the bound 
state peaks ($\Gamma = 100$~MeV~) and a multiplicative factor which restores the 
correct asymptotic $\propto \omega^2$ behavior of the 
continuum part of the spectral functions.

\begin{figure}[h!]
 \begin{minipage}{.45\textwidth}
  \includegraphics[clip,width=\textwidth]{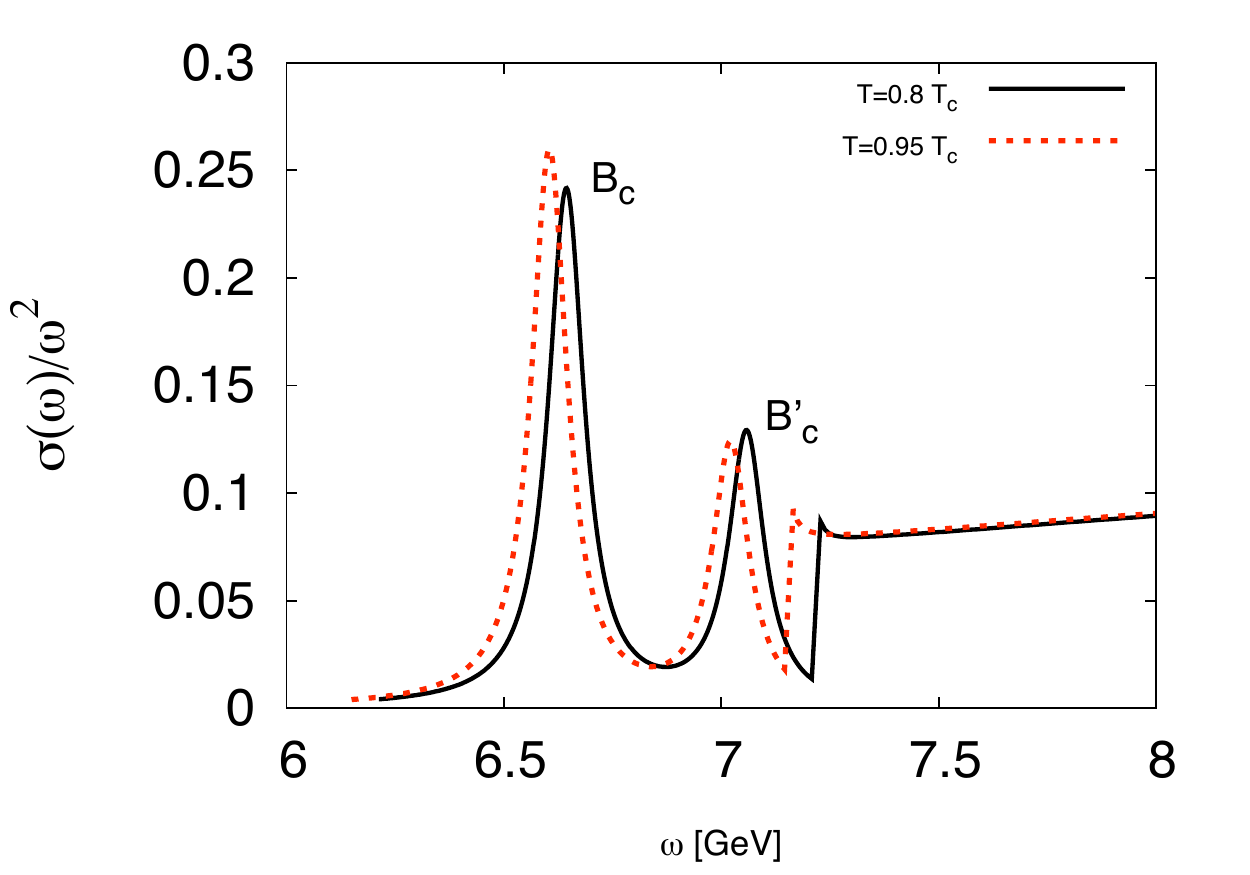}
\caption{The $b\bar{c}$ $S$-wave channel spectral function divided by $\omega^2$ 
as a function of
  $\omega$ at $T=0.8~T_c$ and at $T=0.95~T_c$  }
\label{fig:spfbc1}
 \end{minipage}
 \begin{minipage}{.45\textwidth}
  \includegraphics[clip,width=\textwidth]{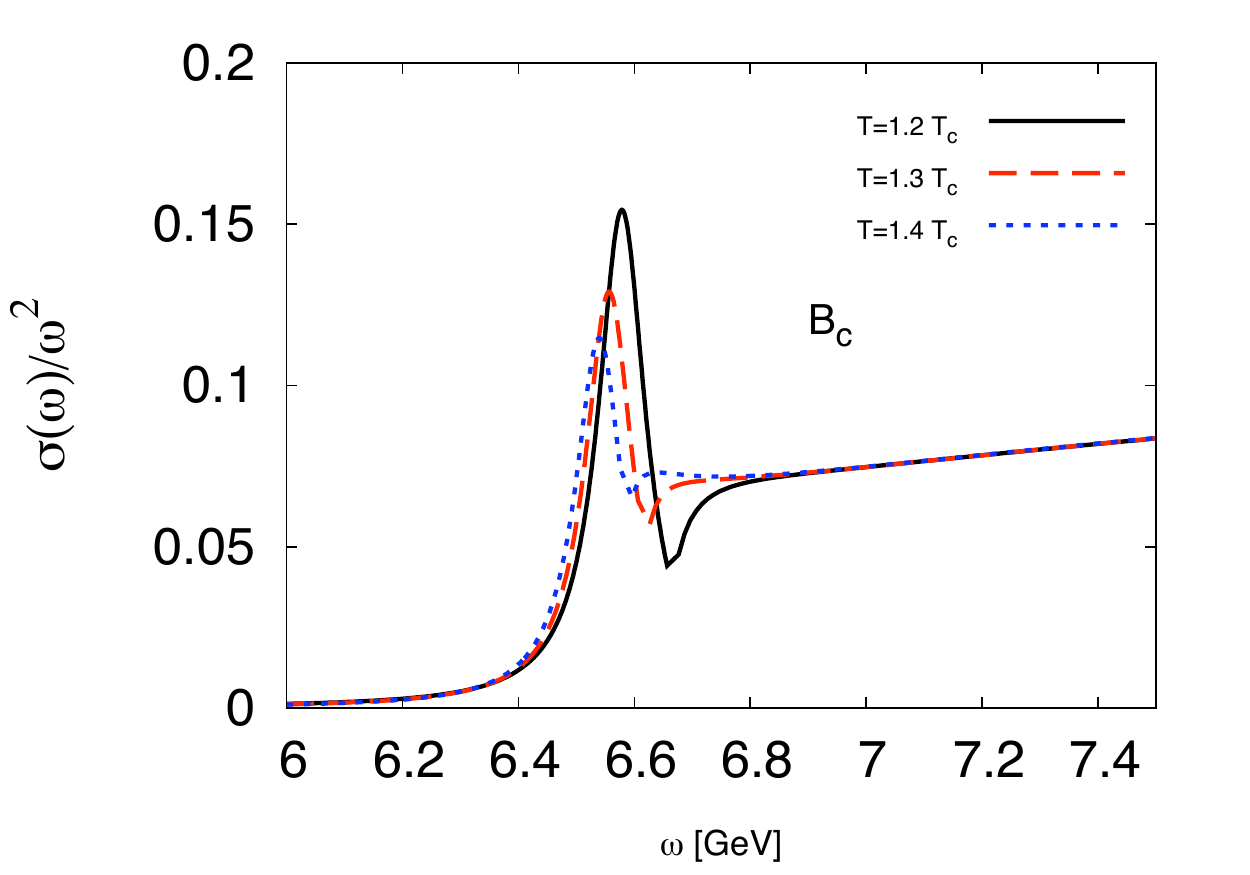}
\caption{ The $b\bar{c}$ $S$-wave channel spectral function divided by 
$\omega^2$ as a function of
  $\omega$ at $T=$~1.2, 1.3 and 1.4~$T_c$}
\label{fig:spfbc2}
 \end{minipage}
\end{figure}

\begin{figure}[h!]
\begin{center}
\includegraphics[clip,width=0.8\textwidth]{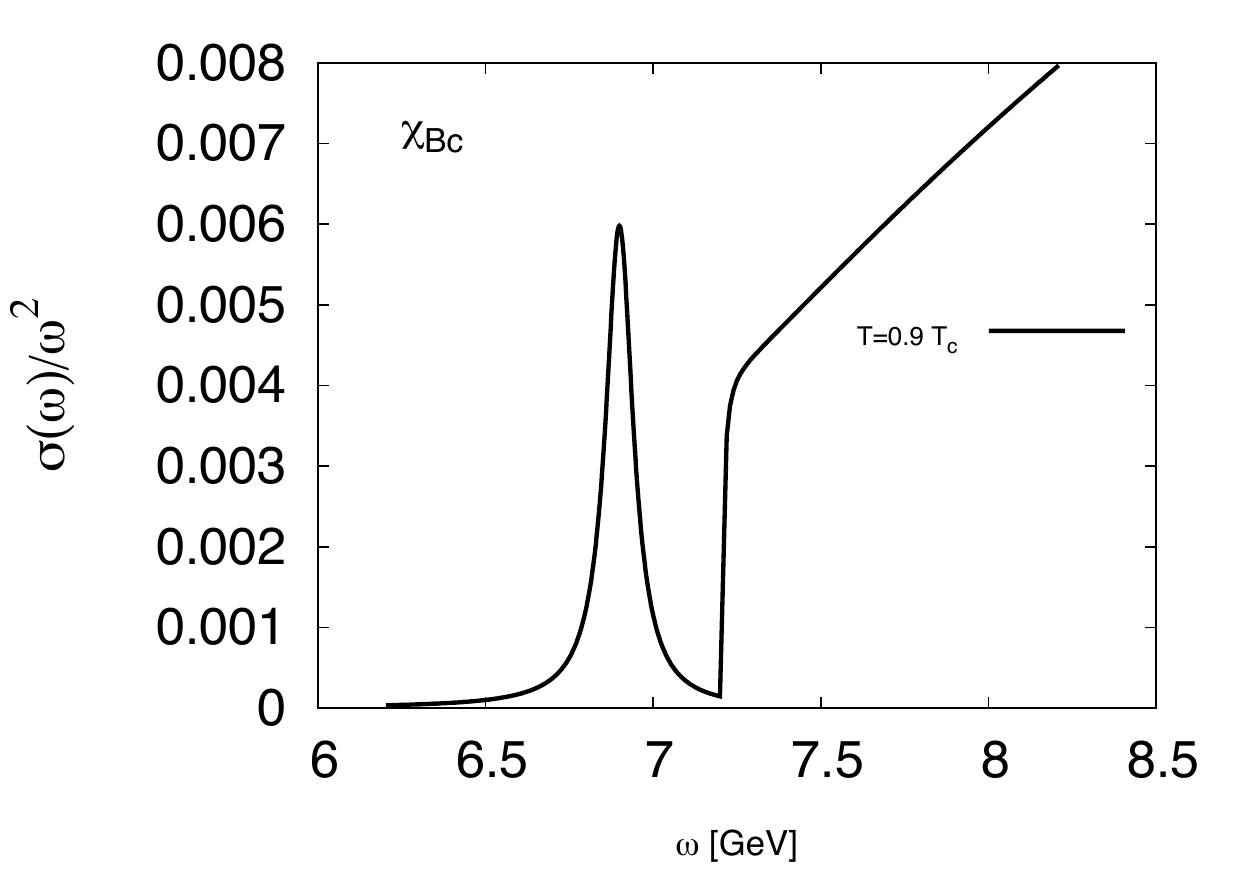}
\caption{ The $b\bar{c}$ $P$-wave channel spectral function divided by 
$\omega^2$ as a function of
  $\omega$ at $T=0.9~T_c$ }
\label{fig:spfchibc}
\end{center}
\end{figure}

Figures \ref{fig:spfbc1} and \ref{fig:spfbc2} show the evolution of the S wave 
spectral function with the temperature. As one can see, 
the fundamental bound state peak survives above critical temperature (up 
to~$\sim~2~T_c$), while the excited state dissociates around~$T_c$.
We also report in Fig. \ref{fig:spfchibc} the shape of the P-wave spectral 
function. The fundamental P-wave state dissociates at $T\sim T_c$.

{\bf Decays}

The $B_c$ lifetime is considerably longer than the one of the other heavy 
quarkonia, due to the lack of direct annihilation decay channels.
Usually the $B_c$ decays are divided into three classes, the first two involving 
the decay of one quark with the other acting as a spectator
and a third one involving annihilation. 
It has already been pointed out \cite{kuma,buch,giri} that an accurate study of 
the $B_c$ decays would allow better estimates of 
some CKM matrix elements, as well as of some leptonic decay constants. We would 
like however to underline the importance of this 
meson as an additional probe for the dynamics of deconfinement in a hot hadronic 
environment.
At present, as already mentioned in the Introduction, a few hundreds decays of 
the $B_c$ into $J/\Psi +\pi$'s has been observed by CMS; 
this is also the best decay channel for the $B_c$ detection in the ALICE 
experiment: the beauty decay  would bring to a 
final state with a $J/\psi$ and a single pion, with a branching ratio estimated 
around $\sim2\%$ \cite{berez}. 

In conclusion we have investigated the survival above the critical temperature 
of a few special quarkonium states, the ones of the 
$B_c$ family, with the main purpose of drawing the attention of the on-going 
experiments at LHC on these intriguing heavy quarkonia. 
As already pointed out for the $J/\Psi$ and $Y$ families, they can survive above 
the temperature for deconfinement of the medium and give
important information on the properties of the hot medium itself.

\section*{Acknowledgments}
One of the authors (P.C.) thanks the Department of Theoretical Physics
 of the Torino University and Istituto Nazionale di Fisica Nucleare, Sezione di 
Torino 
for the warm hospitality in the  final phase of this work.

\end{document}